\documentstyle[pre,aps,epsfig,floats]{revtex}

\begin{document}

\draft

\title{Penetration of dynamic localized states in DC-driven\\
Josephson junction ladders by discrete jumps}

\author{M. V. Fistul and J. B. Page\cite{byline}}
\address{Max Planck Institute for the Physics
of Complex Systems, D-01187, Dresden, Germany}

\date{\today}

\wideabs{ 

\maketitle

\begin{abstract}
We give a theoretical study of unusual resistive (dynamic) localized
states in anisotropic Josephson junction ladders, driven by a DC
current at one edge. These states comprise nonlinearly coupled rotating
Josephson phases in adjacent cells, and with increasing current they are
found to expand into neighboring cells by a sequence of sudden jumps.
We argue that the jumps arise from instabilities in the ladder's
superconducting part, and our analytic expressions for the peculiar
voltage (rotational frequency) ratios and $I$-$V$ curves
are in very good agreement with direct numerical simulations.
\end{abstract}


} 

\section{Introduction}

The subject of large-amplitude anharmonic dynamics in lattices has 
received widespread attention over the past decade. In particular, 
intense theoretical focus has centered on so-called intrinsic localized 
modes (ILMs), also known as discrete breathers, with the result that many of 
their properties are now well understood\cite{ILMs}. These excitations 
result from the interplay between nonlinearity and discreteness, and 
they can be highly localized in perfect lattices, with or without external 
driving. They can occur in a variety of different lattices: recent experiments
have reported vibrational ILMs in a quasi-1D charge-density wave
system\cite{Bishop}, spin-wave ILMs in a quasi-1D antiferromagnetic
system\cite{Sievers}, and discrete breathers in Josephson junction (JJ)
ladders\cite{Ustinov1,Orlando}. 

\begin{figure}[tbh]
\epsfig{file=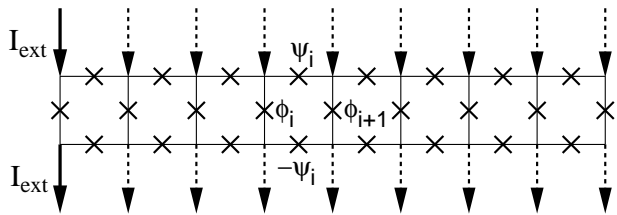,width=8.0cm}
\caption{Sketch of a 1D Josephson junction ladder. Solid arrows: applied
DC edge current. Dashed arrows: bulk current.}
\label{Fig1}
\end{figure}

The latter systems are noteworthy, in that arrays of coupled JJs 
have served for many years as reliable laboratory systems
for studying diverse nonlinear phenomena\cite{StrogLikh}. The
nonlinear dynamics are particularly rich. A single ``small'' JJ 
subject to an applied constant DC bias current can be mapped onto 
the problem of a damped pendulum driven by a constant torque, with the 
dynamical degree of freedom being the Josephson phase 
difference\cite{BandC}. There are thus two qualitatively different 
states, namely a static (superconducting) state and a dynamic (whirling
or resistive) state, with the latter producing a readily measured voltage 
$V \propto \dot \varphi$ across the whirling junction. When several  
junctions are assembled to form a regular array, such as the ladder shown 
in Fig.\ \ref{Fig1}, they become inductively coupled. In the coupled
system, junctions in the superconducting state can also exhibit 
steady-state librations, when JJs in the whirling state are 
present. In view of the mapping onto the pendulum problem, JJ ladders 
share features with lattices of nonlinearly coupled electric dipole 
rotors, driven by an external monochromatic AC electric 
field\cite{BonPage}, but with the important simplification that they can 
be driven with purely DC bias currents. 

Figure \ref{Fig1} sketches an anisotropic JJ ladder consisting of small
JJs of two types, ``horizontal'' and ``vertical,'' which are, 
respectively, perpendicular and parallel to the applied bias current 
(arrows). The anisotropy arises from the different areas of the horizontal 
and vertical junctions and is characterized by the parameter 
$\eta=A_h/A_v=I_c^h/I_c^v$, where $I_c^h$ and $I_c^v$ are the  
the critical currents for each type of junction.
 
References \cite{Ustinov1,Orlando,Ustinov2} report experimental
observations of various discrete breathers in ladders driven by a
{\it homogeneously} applied DC bias current, represented by the dashed
arrows in Fig.\ \ref{Fig1}. For these states, the localized voltage 
patterns have a simple structure involving only two nonzero steady-state 
voltages (rotational frequencies). The breathers 
were found to be stable in the limit of small coupling ($\eta \alt 1$) and 
for bias currents $I_{\rm ext} \alt I_c^v$. For the case of large 2-D 
JJ arrays subject to a homogeneously applied DC bias, more complicated 
inhomogeneous states, with {\it meandering} voltage patterns, have 
also been reported\cite{Misha1}.

Here, we study the dynamics of a JJ ladder with an external
DC bias current applied at only one edge (solid arrows, 
Fig.\ \ref{Fig1}). For increasing bias ($I_{\rm ext} \agt I_c^v$) 
and over a wide range of anisotropies, we find by direct numerical 
simulations that the dynamic state expands into the ladder one cell 
at a time, by a sequence of abrupt jumps. This behavior is in marked 
contrast to the well-known cases of long JJs and JJ parallel arrays 
($\eta~=~\infty$), where the entire system abruptly switches to the 
resistive state at a particular value of the DC bias. It is also
different than the breather case, since all of the junctions within 
the boundary of this localized dynamic state whirl, and at each 
expansion the number of different frequencies (voltages) grows. The 
sequence of $I$-$V$ characteristics and threshold currents can be modeled 
analytically, yielding very good agreement with the numerical results.

\section{Numerical Simulations}

We consider a ladder with a large but finite number of cells $N$. The ladder's 
state is specified by the time-dependent Josephson phases $\{\varphi_i\}$, 
$\{\psi_i\}$, and $\{\tilde \psi_i\}$ for the vertical, upper horizontal
and lower horizontal junctions, respectively, where $i$ denotes the 
cell. We have found in our simulations that the symmetry condition 
${\tilde \psi_i}=-\psi_i$ holds for the phenomena to be discussed here. The 
ladder dynamics are then determined by the coupled nonlinear equations of
motion obtained in Refs.\ \onlinecite{Misha2} and \onlinecite{Grimaldi}:
\begin{eqnarray}
\label{GenEq}
\hat L(\varphi_i)&=&\gamma_i+\frac{1}{\beta_L}[\varphi_{i-1}-2\varphi_i+
\varphi_{i+1}+2(\psi_i-\psi_{i-1})], \nonumber \\
&&i=2,\ldots,N-1, \\
\hat L(\psi_i)&=&\frac{1}{\eta \beta_L}(\varphi_i-\varphi_{i+1}-2\psi_i), \qquad
\quad i=1,\ldots,N, \nonumber
\end{eqnarray}
where the operator $\hat L(\varphi) \equiv \ddot \varphi + \alpha \dot\varphi
+\sin(\varphi)$. The equations for the vertical junctions at $i=1$ and $N$ are
\begin{eqnarray}
\label{BC1}
\hat L(\varphi_1)&=&\gamma_1+\frac{1}{\beta_L}(\varphi_2-\varphi_1+2\psi_1), \\
\hat L(\varphi_N)&=&\gamma_N+
\frac{1}{\beta_L}(\varphi_{N-1}-\varphi_N-2\psi_{N-1}). \nonumber
\end{eqnarray}

Equations (\ref{GenEq}) and (\ref{BC1}) describe each junction within the 
resistively and capacitively shunted junction (RCSJ) model\cite{BandC}, and the 
unit of time is the inverse of the plasma frequency 
$\omega_J \equiv \sqrt{2eI_c/C\hbar}$. Since 
each junction's critical current and capacitance scale with the area, 
$\omega_J$ is independent of the anisotropy parameter $\eta$, as is the 
effective damping constant $\alpha \equiv 1/(\omega_J RC)$. The normalized 
bias current $\gamma_i$ is defined as $I_{i,{\rm ext}}/I_c^v$.  The 
inductive coupling between
junctions is determined by the parameter $\beta_L \equiv 2\pi L I_c^v/\Phi_0$,
where $L$ is the self-inductance of a single cell and $\Phi_0 = hc/2e$ is the 
elementary flux quantum. Coupling beyond that described by $\beta_L$ is
not included. 

We performed direct numerical integration of the equations of motion for 
ladders with $N=20$ cells, using a fifth-order Gear predictor-corrector 
algorithm\cite{AandT}, for a range of anisotropies: $\eta =$ 0.5, 1.0, 2.0, 
3.0, and 5.0. The arrays were underdamped, with $\alpha=0.1$, and we used 
a moderate value of the coupling parameter $\beta_L=0.5$. The external
DC bias was applied at one edge, i.e.\ $\gamma_1=\gamma$ and all other 
$\gamma_i=0$. To simulate the $I$-$V$ curves, we started with all phases 
at zero and gradually increased the external bias $\gamma$ from zero to 50,
in increments
of 0.005. When junctions were present in the whirling state, the MD time 
scale was set by the time-average period of the fastest rotating phase. For 
a given value of gamma, we waited for at least 100 of these reference periods 
before averaging, in order to avoid transients, following which we computed 
the time-averages $\langle \dot \varphi_i \rangle$ and 
$\langle \dot \psi_i\rangle$ over at least
100 additional reference periods. These averages are proportional to the 
average voltages across the junctions. The current was then incremented, with 
the initial phase configuration being that from the preceding MD time step. 
In all runs, the time step was 1/200 of the reference period.

\begin{figure}[tbh]
\epsfig{file=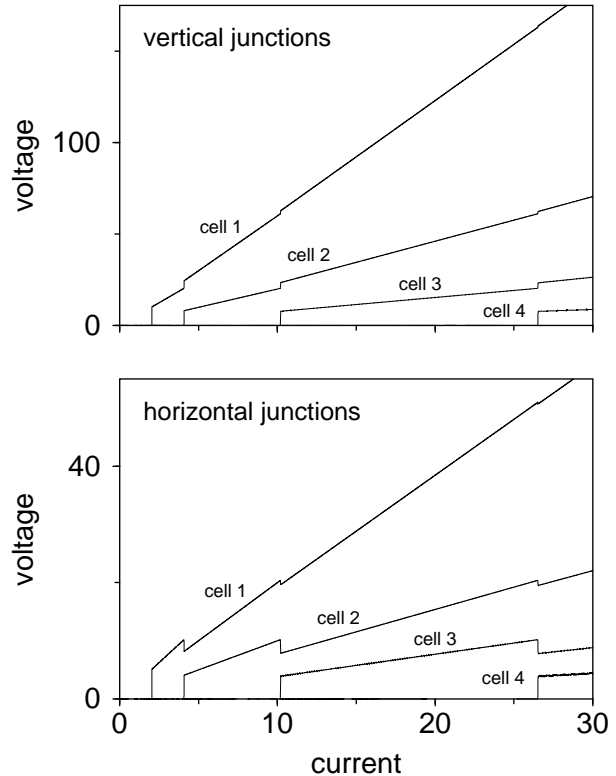,width=8.0cm}
\caption{Numerically simulated $I$-$V_i$ curves for the first four cells
in a 20-cell JJ ladder with anisotropy $\eta=2$. The top and bottom panels
give the time-average frequencies (dimensionless voltages) of the 
Josephson phases for vertical and horizontal junctions, respectively. 
The addition of each new cell to the dynamic state is accompanied by 
abrupt jumps for the other whirling phases. The dimensionless DC bias 
threshold currents for the jumps are $\gamma =$ 2.040, 4.075, 10.20, 
and 26.52.  Extending the current to 50 yielded no additional cells.}
\label{Fig2}
\end{figure}

Our simulated $I$-$V_i$ curves for an anisotropy of $\eta =2$ are shown in 
Fig.\ \ref{Fig2}. The most striking finding is the occurrence of extremely 
sharp voltage jumps. At each of the corresponding threshold currents 
$\gamma^{thr}_n$, a new cell is added to the ladder's dynamic state. 
With the applied current below the first threshold $\gamma^{thr}_1$, all 
junctions are in a static (zero voltage) state. When $\gamma$ exceeds 
$\gamma^{thr}_1$, the first vertical junction and its adjacent top 
and bottom horizontal 
junctions abruptly switch into the dynamic state, with all
other junctions remaining in the zero voltage state---the rotating phases 
are confined to the first cell. As the bias is increased further, all three 
average voltages for this 1-cell dynamic state increase linearly until 
the next threshold current $\gamma^{thr}_2$ is reached, at which point 
the dynamic state suddenly expands into the second cell, accompanied 
by sharp changes of the voltages in the first cell. This process 
continues, yielding successive transitions from $n$-cell dynamic 
localized states to $(n+1)$-cell dynamic states. The distribution of 
threshold currents and voltage ratios depends on the ladder's anisotropy.
Over the range $0<\gamma <50$, the ladder reached a 3-cell state for 
$\eta=0.5$ and 1, a 4-cell state for $\eta=2$, and a 5-cell state 
for $\eta=3$ and 5. In the following, these states will be termed 
$n$-cell edge states.

\begin{figure}[tbh]
\epsfig{file=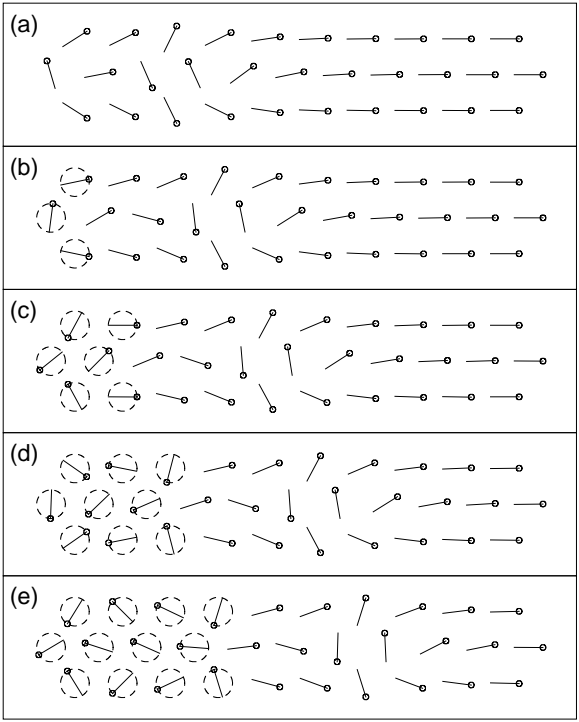,width=8.0cm}
\caption{Instantaneous images of MD-simulated Josephson phase distributions
in the left half of the $\eta=2$ ladder underlying Fig.\ \protect\ref{Fig2},
for different applied edge currents. The dashed circles denote 
junctions in the whirling state. Panel (a): phases immediately before the 
appearance of the 1-cell edge state. Panels (b) through (e): phases just 
after expansion of the edge state into 1, 2, 3, and 4 cells, respectively.
Note the vortex state in the superconducting part of the ladder.}
\label{Fig3}
\end{figure}

An $n$-cell edge state is in striking contrast to an n-cell breather. 
The breather occurs away from the ladder's edge and is homogeneously driven 
by same DC bias current ($I_{\rm ext} \alt I_c$) applied at every cell, 
whereas the edge states are driven by a DC bias ($I_{\rm ext} \agt I_c$) 
applied at just one edge. The edge states have a richer internal 
structure---{\it all} of the junctions within
an edge state are in a nonzero voltage state (see Fig.\ \ref{Fig3}), whereas   
in a breather state, all of the horizontal junctions are in the zero 
voltage state except for those on the breather's 
boundary\cite{Ustinov1,Orlando,Ustinov2}. Moreover, all of a breather's
vertical junction phases rotate at the same average frequency, whereas the 
$n$-cell edge state exhibits a peculiar distribution of average 
frequencies. This frequency (voltage) distribution depends on both $n$ 
and the ladder's anisotropy. For example, our simulations for the $\eta=2$, 
3-cell edge state of Fig.\ \ref{Fig3}(d) yield the ratios given in second 
column of Table \ref{Table1}. Comparison with the third column shows that 
they are in very good agreement with analytic predictions derived below. 

\begin{table}[tbh]
\caption{Average frequency (voltage) ratios for 3-cell edge states. 
The MD ratios are for the $\eta=2$, 3-cell edge state of 
Fig.\ \protect\ref{Fig3}(d), and the predicted ratios were calculated 
from Eqs.\ (\protect\ref{Vert}) and (\protect\ref{Horiz}).}
\begin{tabular}{ccc}
Ratio & MD & Predicted\\
\tableline
$\omega_1^v/\omega_2^v$ & 2.667 &
$(3\eta^2+8\eta+4)/2\eta(1+\eta)=8/3$\\
$\omega_1^v/\omega_3^v$ & 8.006 &
$(3\eta^2+8\eta+4)/\eta^2=8$\\
$\omega_1^h/\omega_2^h$ & 2.499 &
$(\eta^2+6\eta+4)/\eta(2+\eta)=5/2$\\
$\omega_1^h/\omega_3^h$ & 5.017 &
$(\eta^2+6\eta+4)/\eta^2=5$\\
$\omega_3^v/\omega_3^h$ & 2.005 & 2 \\
\end{tabular}
\label{Table1}
\end{table}

The superconducting state forming ahead of the $n$-cell edge state
is also unusual. Fig.\ \ref{Fig3} gives snapshot images of the
Josephson phase distribution for several values of the applied
DC current bias, for the anisotropy $\eta=2$. In panel (a), the
current is just below the first threshold, and one sees a single
{\it Josephson vortex} in the superconducting part of the ladder. The
remaining panels (b)--(e) show the phases just after a new cell
is added to the dynamic state.  At the threshold currents
$\gamma^{thr}_n$, the superconducting state becomes unstable, and the
vortex jumps into the next cell as the edge state expands. Our simulations
show that in general the superconducting state is sensitive to the
anisotropy. Thus for rather small values of $\eta \lesssim 1$, there are
no vortices trapped in the superconducting part of the ladder over the
range of currents studied. For these cases, the Josephson phases of the
vertical junctions in the superconducting portion of the ladder simply 
decrease with distance from the boundary of the resistive portion, 
corresponding to the {\it Meissner state} of the superconductor. With 
increasing anisotropy, single Josephson vortices appear in the 
superconducting portion, as in Fig.\ \ref{Fig3}. 
For large anisotropies ($\eta \sim 5$) more complex {\it vortex trains} are
observed, and we also find that the penetration of the edge state changes the
nature of the superconducting vortex state, rather than simply pushing it
ahead as for $\eta=2$. A detailed discussion of the superconducting
state will be given elsewhere (Ref.\ \cite{FP}).

\section{Theoretical Analysis}

The unusual voltage distributions in the $n$-cell edge states can be 
explained analytically by making use of Kirchhoff's laws, applied
to the time-average currents (normalized to $I^v_c$) and corresponding
dimensionless voltages in each cell. The key assumption is the
coexistence of the resistive and superconducting states in different
portions of the ladder. For a
cell $i$ within an $n$-cell edge state, current conservation gives 
$I^v_i+I^h_i=I^h_{i-1}$, while the voltage condition is 
$I^v_i-\frac{2}{\eta}I^h_i- I^v_{i+1}=0$. Combining these yields an 
equation for just the horizontal currents:
\begin{equation}
\label{DifEq}
I^h_{i+1}+I^h_{i-1}-2\left(1+\frac{1}{\eta}\right)I^h_i=0.
\end{equation}
This equation and the two from which it was derived apply to all cells 
$i$ in $1 \le i \le n$, provided we define $I^h_0 \equiv \gamma,\; 
I^v_{n+1} \equiv 0$, and $I^h_{n+1} \equiv I^h_n$, in order to take 
proper account of the $n$-cell edge state's boundaries. 

Equation (\ref{DifEq}) is readily solved by substituting
$I^h_i = \lambda^i$, which yields two roots, namely $\lambda \equiv 
1+\frac{1}{\eta} + \sqrt{(1+\frac{1}{\eta})^2-1}$ and $1/\lambda$. 
Hence the general solution of Eq.\ (\ref{DifEq}) is 
$I^h_i = c_1 \lambda^i+ c_2 \lambda^{-i}$, where the constants 
$c_1$ and $c_2$ are obtained from the above definitions of 
$I^h_0$ and $I^h_{n+1}$. With the horizontal currents thus determined, 
the vertical currents may be computed from $I^v_i=I^h_{i-1} - I^h_i$. 
The currents are then converted into the average voltages via 
$V^v_i = I^v_i/\alpha$ and $V^h_i=I^h_i/(\alpha \eta)$. The resulting 
voltage distribution within an $n$-cell edge state is ($1\le i\le n$):
\begin{equation}
\label{Vert}
V^v_i=\frac{\gamma(1-\lambda)(\lambda^{i-1} -\lambda^{2n+1-i})}
{\alpha(\lambda^{2n+1}+1)},
\end{equation}
\begin{equation}
\label{Horiz}
V^h_i=\frac{\gamma (\lambda^{i} +\lambda^{2n+1-i})}
{\alpha \eta (\lambda^{2n+1}+1)}.
\end{equation}

Equations (\ref{Vert}) and (\ref{Horiz}) give the predicted voltage 
ratios in Table \ref{Table1}, which for $\eta=2$ are seen to be 
in very good agreement with our MD simulations. Indeed, we find 
that for all of the values of $\eta$ studied, the predicted $I$-$V$ 
curves are in very good agreement with the MD curves, such as those
of Fig.\ \ref{Fig2}. Only the values of the current thresholds 
for the jumps are left undetermined by these equations.  

\begin{figure}[tbh]
\epsfig{file=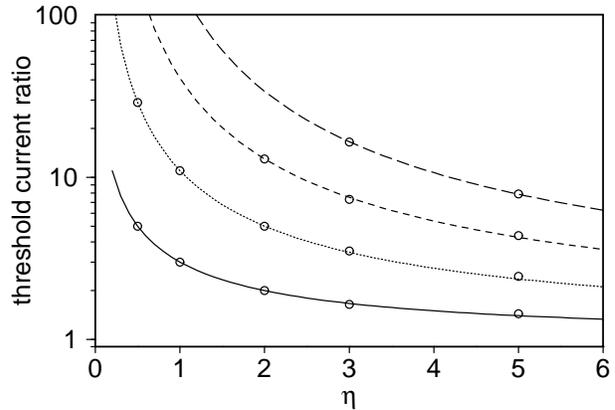,width=8.0cm}
\caption{Threshold current ratios $\gamma^{thr}_n/I_{dp}$.  Curves:
predicted ratios vs.\ anisotropy $\eta$, from Eq.\ (\protect\ref{Ithr}).
Starting at the bottom, the curves are for $n =$ 2, 3, 4 and 5.
Circles: numerical ratios from the MD simulations.}
\label{Fig4}
\end{figure}

We can also predict the distribution $\{\gamma^{thr}_n\}$ of 
threshold currents for each $\eta$ by assuming that the superconducting 
state associated with the $(n-1)$-cell 
edge state becomes unstable and converts to the $n$-cell edge state when 
the current $I^h_n$ reaches a depinning current $I_{dp}$, which we take
to be independent of $n$. This yields an expression for the threshold 
currents 
\begin{equation}
\label{Ithr}
\gamma^{thr}_n =
I_{dp}\frac{\cosh{[(n-\frac{1}{2}) \ln \lambda]}}
{\cosh{(\frac{1}{2} \ln \lambda)}}.
\end{equation}
The ratios $\gamma^{thr}_n/I_{dp}$ predicted by Eq.\ (\ref{Ithr}) for
$n=$ 2, 3, 4 and 5 are plotted versus $\eta$ in Fig.\ \ref{Fig4}. To
compare with the MD results, we fit $I_{dp}$ to the 
first observed MD threshold current for each $\eta$, namely 
$\gamma^{thr}_1 =$ 1.295, 1.510, 2.040, 2.525, and 4.010, for 
$\eta =$ 0.5, 1.0, 2.0, 3.0, and 5.0, respectively. The resulting 
MD ratios are shown by the circles in Fig.\ \ref{Fig4} and agree well
with the predictions. 

\section{Hysteresis}

\begin{figure}[tbh]
\epsfig{file=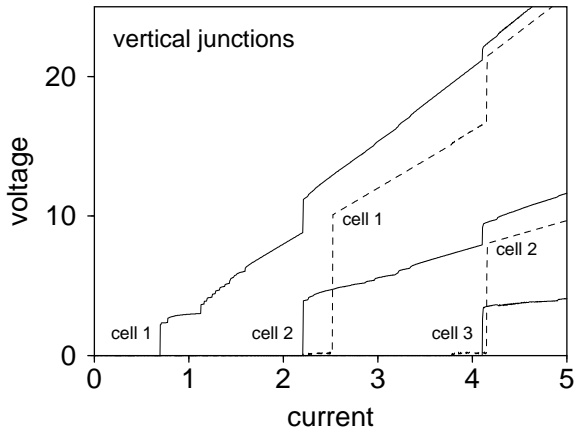,width=8.0cm}
\caption{Hysteresis. Shown are the numerically simulated $I$-$V$ curves for
the vertical junctions of the first few cells in a 20-cell JJ ladder with
anisotropy $\eta = 3$. The bias current was first increased (dashed curves)
from zero to 50 in increments of 0.005, resulting in a 5-cell edge state,
at which point the current was decreased (solid curves) back to zero with
the same increment. At each value of the current, we waited 400 reference
periods and then averaged the voltages over an additional 200 periods.
Only a small current range is shown, in order to illustrate clearly the
differences between the increasing- and decreasing-current cases. The
latter case has a different sequences of threshold currents and
exhibits nonlinear regions and small steps not associated
with the disappearance of the edge states.}
\label{Fig5}
\end{figure}

Despite their rich structure of frequency ratios, the $n$-cell edge states
are found to be highly stable in our simulations, for the case of increasing
current. However, we also find notable hysteresis effects when the
simulations are started with an $n$-cell edge state for large $n$ and
the applied DC current is gradually {\it decreased} to zero. Figure
\ref{Fig5} is representative of the hysteretic behavior encountered.
The threshold currents and stability properties for the sequence of
down-conversions $\{n \rightarrow n-1\}$ are very different than for
the increasing-current case. In particular, we observe resonant
steps, switching processes, and nonlinear regions in the $I$-$V$ curves.
We believe that all of these features arise from the resonant interaction
between the n-cell edge states and other excitations, both localized and
delocalized, as will be discussed elsewhere\cite{FP}.

\section{Summary}

In summary, our numerical simulations have revealed unusual localized 
dynamic states in anisotropic JJ ladders subject to a DC bias current 
at one edge. Increasing the bias causes these states to expand by adding 
single cells in a sequence of sudden jumps, giving rise to a diverse set 
of voltage distributions and sharp changes in the $I$-$V$ curves. 
This behavior occurs for a wide range of parameters and 
should be observable through the $I$-$V$ characteristics or by direct
visualization using low temperature scanning laser microscopy 
techniques\cite{Ustinov1,Ustinov2,Misha1}.

\acknowledgements

We thank A. V. Ustinov and S. Flach for useful discussions. J. B. Page 
gratefully acknowledges the Max Planck Institute for the Physics of 
Complex Systems, Dresden, for their support and hospitality.

\end{document}